\documentstyle[11pt]{article}   

\begin{document}

%%%%%%%%%%%%%%%%%% Reference Defs %%%%%%%%%%%%%%%%%%
\def\NPB#1#2#3{Nucl.\ Phys.\ {\bf B#1}, #3 (19#2)}
\def\NPPS#1#2#3{Nucl.\ Phys.\ Proc.\ Suppl.\ {\bf #1}, #3 (19#2)}
\def\PLB#1#2#3{Phys.\ Lett.\ {\bf B#1}, #3 (19#2)}
\def\PLBold#1#2#3{Phys.\ Lett.\ {\bf#1B}, #3 (19#2)}
\def\PRD#1#2#3{Phys.\ Rev.\ {\bf D#1}, #3 (19#2)}
\def\PRL#1#2#3{Phys.\ Rev.\ Lett.\ {\bf#1}, #3 (19#2)}
\def\PRep#1#2#3{Phys.\ Rep\. {\bf#1}, #3 (19#2)}
\def\ARAA#1#2#3{Ann.\ Rev.\ Astron. Astrophys.\ {\bf#1}, #3 (19#2)}
\def\ARNP#1#2#3{Ann.\ Rev.\ Nucl.\ Part.\ Sci.\ {\bf#1}, #3 (19#2)}
\def\IJMPA#1#2#3{Int.\ J.\ Mod.\ Phys.\ {\bf A#1}, #3 (19#2)}
\def\MPLA#1#2#3{Mod.\ Phys.\ Lett.\ {\bf A#1}, #3 (19#2)}
\def\ZPC#1#2#3{Zeit.\ f\"ur Physik {\bf C#1}, #3 (19#2)}
\def\APJ#1#2#3{Ap.\ J.\ {\bf #1}, #3 (19#2)}
\def\AP#1#2#3{{Ann.\ Phys.\ } {\bf #1}, #3 (19#2)}
\def\RMP#1#2#3{{Rev.\ Mod.\ Phys.\ } {\bf #1}, #3 (19#2)}
\def\CMP#1#2#3{{Comm.\ Math.\ Phys.\ } {\bf #1}, #3 (19#2)}
\def\EPJC#1#2#3{Eur.\ Phys.\ J.\ {\bf C#1}, #3 (19#2)}
\def\PTP#1#2#3{Prog.\ Theor.\ Phys.\ {\bf#1}, #3 (19#2)}
%%%%%%%%%%%%%%%%%%%%%%%%%%%%%%%%%%%%%%%%%%%%%%%%%%%%%%%%%%%%%%
%Incorporating the figure
\newcommand{\postscript}[2]
{\setlength{\epsfxsize}{#2\hsize}
\centerline{\epsfbox{#1}}}
\newcommand{\newc}{\newcommand}
\newcommand{\gsim}{ \mathop{}_{\textstyle \sim}^{\textstyle >} }
\newcommand{\lsim}{ \mathop{}_{\textstyle \sim}^{\textstyle <} }
\newcommand{\tev}{\,{\rm TeV}}
\newc{\gev}{\,{\rm GeV}}
\newc{\mpl}{M_P}
\newc{\mz}{m_Z}
\newc{\mw}{m_W}
\newc{\mx}{m_{GUT}}
\newc{\mn}{M_d}
\newc{\mr}{M_R}
\newc{\md}{\mn}
\newc{\mstr}{M_s}
\newc{\mstrn}{\mstr}
\newc{\beq}{\begin{equation}}
\newc{\eeq}{\end{equation}}
\newc{\bea}{\begin{eqnarray}}
\newc{\eea}{\end{eqnarray}}
\newc{\ie}{{\it i.e.\/}}
\newc{\eg}{{\it e.g.\/}}
\newc{\tr}{\mbox{Tr}\,}
\newc{\trp}{\mbox{Tr}'\,}
\renewcommand{\bar}{\overline}
\def\vev#1{\left\langle #1 \right\rangle}
%%%%%%%%%%%%%%%%%%%%%%%%%%%%%%%%%%%%%%%%%%%%%%%%%%%%%%%%%%%%%%
\catcode`@=11
% Redefine caption to put text and formulas in smaller font
\long\def\@caption#1[#2]#3{\par\addcontentsline{\csname
  ext@#1\endcsname}{#1}{\protect\numberline{\csname
  the#1\endcsname}{\ignorespaces #2}}\begingroup
    \small
    \@parboxrestore
    \@makecaption{\csname fnum@#1\endcsname}{\ignorespaces #3}\par
  \endgroup}
\catcode`@=12
%%%%%%%%%%%%%%%%%%%%%%%%%%%%%%%%%%%%%%%%%%%%%%%%%%%%%%%%%%%%

\begin{titlepage}
\begin{flushright}
{\rm
LBNL--42577\\
LANCS--TH/9822\\
hep-ph/9812234\\
December 1998
}
\end{flushright}
\vskip 1.5cm

\begin{center}
%%%\title
{\Large \bf
D-term Inflation and M-theory \\
}
\vskip 0.7cm
{\large
Christopher Kolda$\,{}^\dag$ and David H.~Lyth$\,{}^\ddag$}
\vskip 0.4cm
{\em
${}^\dag$ Theoretical Physics Group, 
  Lawrence Berkeley National Laboratory, \\ 
  University of California, Berkeley, CA 94720, USA\\[2mm]
${}^\ddag$ Department of Physics, Lancaster University, Lancaster, 
  LA1~4YB, UK
}
\end{center}
\vskip .5cm

%%%\date{October 1998}
%%%\maketitle

\begin{abstract}

Models of supersymmetric $D$-term inflation require a new mass scale
near $10^{15-16}\gev$ in order to match the density perturbation
spectrum observed by COBE. Attempts to obtain such a scale from the 
anomalous U(1) of string theories fail in most string models. However
there is hope that models based on non-standard embeddings in M-theory 
can solve the discrepancy. We will show that such models still
suffer from the other drawback 
of $D$-inflationary models, namely that Planckian field values are required
to drive inflation. Thus it is hard to understand why the inflaton
potential remains so flat without imposing stringent symmetries on the 
superpotential. We 
also examine a fascinating quasi-fixed point behavior for the gauge
coupling of the anomalous U(1) in these extra-dimensional models, and
show that either the presence of large numbers of fields in the 5-dimensional
bulk or strongly suppressed U(1) gauge couplings
is required in order to restore naturality to the inflationary
potential.  

\end{abstract}

\end{titlepage}
\setcounter{footnote}{0}
\setcounter{page}{1}

%\newpage

\section{Introduction}

Particle physics models of inflation must
incorporate two key
elements: a potential for the inflaton that is sufficiently flat to
allow a long period of inflation ($\gsim40$ to 60 $e$-folds) and a mechanism 
by which that inflation is ended. Hydrid, or ``two-field,'' inflationary
models~\cite{linde,lr} 
are probably the simplest and most attractive solution to both
of these problems: while one field (the inflaton) slowly rolls, 
inflation is driven by the false vacuum
energy of the second field which is prevented from dropping to its
true minimum by its coupling to the inflaton. Eventually, the inflaton 
vacuum expectation value falls below a critical value at which
it can no longer support the second field in its false vacuum. At that 
time, the second field drops smoothly to its true (zero energy) 
minimum and inflation ends.

Within the context of supersymmetry (SUSY), a large number of hybrid
inflationary models have been built which, at lowest order, provide
all of the elements necessary for successful inflation. These SUSY
models fall into two classes, because (in the language of global
SUSY) vacuum energy is always either the result of non-zero $F$-terms 
or $D$-terms. The models of $F$-term inflation are
known to suffer from a serious problem. If $\psi$ is the field whose
$F$-vev drives inflation ($|F_\psi|^2=3\mpl^2 H^2$ where $H$ is the Hubble
parameter) and $\phi$ is the
inflaton which is slowly rolling down its potential, then at the
non-renormalizable level, the K\"ahler potential generically contains
terms of the form:
\beq
K=\frac{1}{\mpl^2}\,\phi^\dagger\phi\psi^\dagger\psi
\label{F}
\eeq
where $\mpl=(8\pi G_N)^{-1/2}$.
In a full supergravity scenario, terms such as this one, along with
others that arise in the scalar potential, produce a mass term for
the inflaton $m_\phi^2=a|F_\psi|^2/\mpl^2=aH^2$ with $a\sim{\cal
O}(1)$. The resulting
equation of motion for $\phi$ is not slow-rolling, but near
critical damping: $\ddot\phi+3H\dot\phi+2aH^2\phi=0$. This feedback of 
the SUSY-breaking $F$-type vacuum energy into the inflaton potential
is very general~\cite{finfl}, though ways of avoiding this problem have been
discussed in Ref.~\cite{lr}.

The alternative class of models, those in which $D$-terms dominate the 
vacuum energy~\cite{dinfl}, do not suffer from this same malady since terms
corresponding to Eq.~(\ref{F}) do not exist unless the inflaton
carries a gauge charge, which one would already assume not to be the
case. Models of $D$-term inflation make use of a U(1)
gauge symmetry for which one can write an explicit energy density
which is both SUSY- and gauge-invariant: the Fayet-Iliopoulos term,
$g\xi\int\! d^4\theta\,V$. In the superpotential, $\phi$ couples to two
fields, $\psi$ and $\bar\psi$, which carry U(1) charges
$\pm1$ respectively: $W=\lambda\phi\bar\psi\psi$. The resulting
potential is:
\beq
V=|\lambda\phi|^2\left(|\psi|^2+|\bar\psi|^2\right) +|\lambda
\bar\psi\psi|^2+\frac{g^2}{2}\left(|\psi|^2-|\bar\psi|^2
+\xi\right)^2 
\label{Vtree}
\eeq
(We will define
$\xi>0$ for convenience.) For large initial values of 
$\phi$, the $\bar\psi$ and $\psi$ fields
receive large masses, forcing them to $\bar\psi=\psi=0$. The resulting 
vacuum energy is then $V_0=g^2\xi^2/2$, driving a period of inflation and
breaking SUSY. With SUSY broken, a potential for $\phi$ is generated at 
one-loop:
\beq
V(\phi;\mu)=\frac12g^2\xi^2\left[1+C\frac{g^2}{8\pi^2}\,\log\left(
\frac{|\lambda\phi|}{\mu}\right)
\right]\ ,
\label{Vloop}
\eeq
where $C\geq1$ is the number of pairs of $\bar\psi,\psi$ fields to which $\phi$
couples with strenth $\lambda$, and 
$V$ is to be evaluated at the scale $\mu$.
Ideally, $\phi$ rolls slowly in its logarithmic potential until it
reaches it critical value, $\phi_c=g\sqrt\xi/\lambda$, at which time 
$\bar\psi$ falls quickly to $\sqrt{\xi}$, SUSY is restored and
inflation ends. 

This model has two problems, the first well-known but model-dependent, 
the second less-known but generic. The first problem stems from a
calculation of the density perturbations resulting from the above
potential. COBE imposes the normalization~ (see, \eg, Ref.~\cite{lr}):
\beq
(V_0/\epsilon)^{1/4}=6.7\times10^{16}\gev
\label{cobe}
\eeq
where $\epsilon$ is one of the slow-roll parameters:
$\epsilon\equiv\frac{1}{2}\mpl^2(V'/V)^2$. In principle, this expression should
be evaluated $N$ $e$-folds before the end of inflation, where $N$ is the number
of $e$-folds remaining after the scales probed by COBE left the horizon
$(40\lsim N\lsim60)$. Plugging into these the
potential $V(\phi;\mu=\phi)$ 
from above (the result is actually almost independent
of the details of $V(\phi)$), one finds
\beq
\sqrt\xi=8.5\times10^{15}\gev\,\times\,\left(\frac{50C}{N}\right)^{1/4}\ 
. 
\eeq

In and of itself, this is not a
problem, since we have not specified a source for $\xi$ in the toy
model above. However, the most attractive mechanism for generating $\xi$
of this order is via the pseudo-anomalous U(1) of string theory. In
many string theories, one finds a U(1) gauge group, which at the level 
of the massless fermion fields, appears to be anomalous (it has both
[U(1)]${}^3$ and ${\cal G}\times$[U(1)]${}^2$ anomalies for each group 
${\cal G}$ in the four-dimensional theory). The anomaly is fictitous,
however, as it is eliminated by the Green-Schwarz mechanism:
the model-independent axion transforms under the U(1), cancelling the 
$F\widetilde F$-terms generated by the anomaly in the fermion sector so 
that the action remains invariant. However, the
transformation of the axion field generates a Fayet-Iliopoulos term
for the U(1) which can be calculated in any string model. In
heterotic models, the result was first obtained in Refs.~\cite{dsw}:
\beq
\xi=\frac{1}{192\pi^2}(\tr Q)\,\mstr^2
\label{xi}
\eeq
where Tr$\,Q$ sums the U(1) charges of all massless states, 
and $\mstr$ is the string scale, equal to $g\mpl$ at weak coupling.
For phenomenological values
of $g\approx0.7$, $\xi$ is always larger than $(2\times10^{16}\gev)^2$; in
realistic string models, Tr$\,Q\sim100$ usually, so that typical $\xi$ 
are approximately (few$\times10^{17}\gev)^2$. In either case, the natural 
scale for $\xi$ greatly exceeds the bound imposed by COBE~\cite{lrd}.

We now know that Eq.~(\ref{xi}), with $\mstr=g\mpl$, holds more generally
in string theory~\cite{jmr}, as long as the axion whose transformation cancels
the anomaly is the partner of the perturbative dilaton. 
Recently, however, string
theories have been discussed in which the anomalous U(1) and its Green-Schwarz
axion  arise non-perturbatively. Such models would avoid
the constraint of Eq.~(\ref{xi}). In this paper we will examine extensions of
one particular string model, the so-called Ho\v rava-Witten (HW)
model~\cite{hw}. The minimal HW model itself contains no anomalous U(1), but
its generalizations can~\cite{bddp,lpt}.

The second problem faced by models of $D$-term inflation is more
generic. Up until now, we have assumed that inflation ended when
$\phi=\phi_c$. However, inflation also requires slow-roll, so that
$\epsilon$ defined above must be $\,\ll1$. 
Likewise, inflation also requires that
the second slow-roll parameter, $\eta\equiv\mpl^2|V''/V|$, 
be $\,\ll1$, since $\eta$ measures the consistency of the slow-roll
approximation. Thus inflation could end
for $\phi\gg\phi_c$ if either $\epsilon\sim1$ or $\eta\sim1$ first. This is
in fact what happens~\cite{km}. Using the potential in Eq.~(\ref{Vloop}), 
\beq
\eta(\phi)=\sqrt{\frac{C\alpha}{2\pi}}\,\frac{\mpl}{\phi}
\label{eta}
\eeq
where $\alpha=g^2/4\pi$ as usual. Inflation will end when $\eta\sim1$, which
happens when $\phi=\phi_f\simeq\sqrt{C\alpha/2\pi}\mpl$. For the
phenomenological
value of $\alpha\simeq1/25$ and $C=1$, inflation ends for
$\phi_f\simeq0.1\mpl$,
which is much larger than $\phi_c$ given the COBE normalization of $\xi$ and
$\lambda\sim{\cal O}(1)$.

This value in and of itself is not overly large, but we must remember
$\phi_f$ represents the {\em end}\/ of inflation. To calculate the initial
value of $\phi$ needed for $N$ $e$-folds, we use the relation
$dN=-(V/V'\mpl^2)\,d\phi$. Solving, one finds
\beq
\phi_i\simeq\sqrt{\frac{C\alpha N}{\pi}}\mpl
\eeq
which means $\phi_i\simeq0.8\mpl$ for $N=50$.
As in models of chaotic inflation, field values so close to the Planck scale
are not necessarily unreasonable. However, they do mean that
the field theory may not be
under control. In particular, corrections to the  
superpotential of the form $\phi^n/\mpl^{n-3}$ 
are not particularly suppressed, so that the
slope of $V(\phi)$ is naturally ${\cal O}(1)$. It is possible to eliminate the
higher-order terms with $R$-symmetries, for example, though this may
have a strong impact on the post-inflationary reheating~\cite{km}. In
any case, one is still left
with the conclusion that the inflationary dynamics is occurring at
scales where string theory or quantum gravity replace field theory.
Thus the model may not be internally consistent.

In this paper, we examine the solution to the first problem (COBE
normalization) which is provided by the extended HW models and show that the
second problem (Planck-scale inflation) is still present. Thus we will show
that models of the HW-type do not solve {\em all}\/ 
the problems faced by $D$-term inflation.

\section{Extra Dimensions and Anomalous U(1)'s}

The Ho\v rava-Witten action~\cite{hw} 
represents one particular limit of M-theory in
which 11-dimensional supergravity is compactified in two stages:
first, dimensions $6\ldots11$ are compactified on a 6-dimensional
Calabi-Yau manifold at the
string scale as is customary; but the fifth dimension is compactified on a line
segment (${\bf S}^1/{\bf Z}_2$) 
whose length, $\pi R$, is much larger than the inverse string scale.
After the compactification of the
fifth dimension, the theory resembles the usual $E_8\times E_8$ heterotic
string but with the surviving subgroups of each $E_8$ (one $E_8$ for the
visible sector and one for the hidden
sector) on  two different 4-dimensional ``walls,'' a distance $\pi R$ apart.
Standard Model gauge interactions and charged matter are confined to
one or the other wall and so the Standard Model 
gauge theory appears 4-dimensional even at distances shorter 
than $R$. Gravity, however, knows
about the additional dimension immediately at $R$. 

Once gravitational interactions become 5-dimensional, 
the gravitational potential falls off more quickly. Matching the 4- and
5-dimensional
theories at $R$, one finds:
\beq
\mpl^2=(2\pi R)M_5^3=(2\pi R)V_{CY}M_{11}^9
\label{M5}
\eeq
where $\mpl$ is the usual 4-dimensional (reduced) Planck mass, 
$M_5 (M_{11})$ is its equivalent in the 5(11)-dimensional theory, and 
$V_{CY}$ is the volume of the 6-dimensional Calabi-Yau manifold on which the
full 11-dimensional theory is compactified down to $d=5$; we identify the 
string scale by $\mstrn=V_{CY}^{-1/6}$. Witten proposed that this
matching could explain why the gauge couplings of the MSSM do not meet at the
usual weak-coupling string  scale~\cite{gcu} 
but rather at $3\times10^{16}\gev$. In this theory, the
running of the 4-dimensional gauge couplings is left unchanged, but for 
$\mr\equiv1/\pi R\simeq5\times10^{15}\gev$, one finds that the 
``true'' 11-dimensional Planck mass, $M_{11}$, 
moves down to roughly $6\times10^{16}\gev$ and the corresponding
string scale, $\mstrn$, to $3\times10^{16}\gev$~\cite{gcu,bd,lpt}. 
Notice that both $\mpl$ and $M_5$ are larger than the true $d=11$ Planck
scale.

The minimal HW model contains no anomalous U(1) gauge groups. Such
a group would have to interact simultaneously with fields on both walls and
must then live in the intermediate ``bulk,'' the region between the
two 4-dimensional walls. The action, however, contains no fields
which can play the role of the new photon. Thus, this model appears to
solve neither of the two problems present in $D$-term inflation which we
discussed earlier. 

Nonetheless, the minimal HW action can be extended in 
a way which {\em is}\/ interesting to us. 
Non-perturbative effects (represented perhaps by
adding additional D-branes to the model) can give rise to gauge fields 
and charged matter in the bulk~\cite{bddp,lpt}. 
One of these
groups could be an anomalous U(1), for which a Green-Schwarz mechanism
would still operate. Assuming that such a U(1) does arise, the resulting
Fayet-Iliopoulos term can be calculated~\cite{bddp}. The result reproduces
exactly Eq.~(\ref{xi}).
where now the trace $\tr Q$ is over all matter in the bulk and on either wall
and $\mstrn$ is henceforth the true string scale, \ie, 
the unification scale of the MSSM gauge couplings; 
that is, the {\em apparent}\/ 4-dimensional string scale has been replaced with
the true 11-dimensional one. Given $\mstrn\simeq3\times10^{16}\gev$, it
appears that deriving values of $\xi$ consistent with the COBE bound is no
longer a problem, as was noticed in Ref.~\cite{bddp}. 
This then would appear to be a great success for both M-theory and $D$-term
inflation.

\section{Slow-Roll and the End of Inflation}

Before examining $D$-term inflation in the context of these
generalized HW models, 
we need to verify that the usual inflation model-building will
still work. In that vein, we would like to demand
that $H<\mr$
during inflation (\ie, the Hubble radius is larger than $R$ so that on
cosmological scales the universe can be described by a $d=4$ metric)
and that inflation is driven by $d=4$
particle dynamics, so that the usual calculation of density perturbations
holds~\cite{lyth}.

We can actually check to see if $H<\mr$ self-consistently in these schemes.
Given the COBE normalization in Eq.~(\ref{cobe}) (with $\epsilon<1$):
\beq
H=\left(\frac{V_0}{3\mpl^2}\right)^{1/2}<1.1\times10^{15}
\label{hmr}
\eeq
which is obviously smaller than $\mr\simeq5\times10^{15}\gev$.
(This result makes more precise a result first presented in Ref.~\cite{lr}.)
Thus we
can safely do cosmology in 4 dimensions regardless of the existence
of the bulk.

Given $H<\mr$, the inflationary potential is unchanged from
Eqs.~(\ref{Vtree}) and (\ref{Vloop}), though now
$\xi\sim10^{15-16}\gev$, consistent with COBE. The 
slow-roll parameter $\eta$ is still given by Eq.~(\ref{eta}) and the arguments
proceed as before. In particular, the requirement of $N$ $e$-folds
of inflation gives $\phi_i>\sqrt{C\alpha N/\pi}\mpl$ again.
But there is a further requirement which must hold in order for the theory to
remain in the field-theoretic regime, and for the non-renormalizable
contributions to $W$ to be naturally suppressed,
namely $\phi\ll M_{11}$ (recall that the
field theory cut-off is no longer $\mpl$, but is instead $M_{11}$, and so all
non-renormalizable operators are suppressed only by $M_{11}$).
Combining these two requirements yields:
\beq
M_{11}\gg\sqrt{\frac{C\alpha N}{\pi}}\mpl\simeq 4\sqrt{\frac{C\alpha N}{50}}\mpl\ .
\label{self}
\eeq
For $M_{11}\simeq6\times10^{16}\gev$, the above constraint requires
a very small value for $\alpha$, namely $\alpha\lsim 10^{-5}$.

We turn now to the question of the natural size of $\alpha$, the gauge coupling
of the anomalous U(1), evaluated at the string scale. Even without specifying
a particular realization of the anomalous U(1) in 
the bulk, we can still broadly consider the expectations for
$\alpha(\mstrn)$. If $\alpha$ is roughly the same size as the unified coupling
on the Standard Model wall, then $\alpha\sim1/25$, clearly too
large. It has also been found~\cite{lpt} in particular embeddings that $\alpha$
is intermediate to the gauge couplings on the two walls; under the
usual assumption that the theory on ``our'' wall is the weaker, then
again $\alpha$ is again too large. We cannot rule out the other
possibility, that our wall is the strongly-coupled theory, which means
that $\alpha$ could be very small indeed (though we know of no
suggestions to make it as small as $10^{-5}$).

Of course, the value of $\alpha$ being discussed is its value as
measured on the 4-dimensional wall of the Standard Model at the energy
scale of the inflationary potential, $\xi$. This value can be
calculated from its value at the 5-dimensional string scale via the
usual renormalization group procedure.
How $\alpha$ will run depends intimately on whether or not there
exist any charged fields living in the bulk. If there are no charged fields in
the bulk, then $\alpha$ as measured on the wall containing the inflationary
potential will run only according to the usual 4-dimensional $\beta$-function:
\beq
\alpha^{-1}(\mu)=\alpha^{-1}(\mstrn)+\frac{1}{2\pi}(\tr Q^2)\log\left(
\frac{\mstrn}{\mu}\right)\ 
\eeq
at any scale $\mu<\mstrn$. Here $\tr Q^2$ sums over all the fields 
with mass $m<\mu$. The resulting change in $\alpha$ is running down
from the string scale is therefore insignificant and does not rescue inflation. 

However, in the more likely case that there 
{\em are}\/ charged fields in the bulk, the
result changes significantly. Now, the gauge coupling of the U(1) runs not in 4
dimensions but in 5. Or equivalently, the gauge coupling runs in 4 dimensions,
but with the contributions of the bulk Kaluza-Klein (KK) excitations 
included. The renormalization group equation is simplest to derive from this
latter point of view~\cite{ddg}. 
To do so, imagine running $\alpha$ from $\mr$ to some arbitrary
scale $\mu$ such that $\mr\leq\mu\leq\mstrn$. Each state in the ``massless''
spectrum (by which we means states
with mass $\ll\mr$, either on the walls or in the bulk) contributes to the
$\beta$-function a factor $Q^2/2\pi$. But for 
each massless state in the bulk, there is a tower of KK excitations with masses
$m_n^2=n^2\mr^2$ where $n=1\cdots\infty$. At the scale $\mu$ there are
$\mu/\mr$ states with masses $m<\mu$. Unless $\mu$ is very close to $\mr$,
these KK contributions will overwhelm the contributions from the massless
states in the bulk and on the walls, and so we can drop these latter
contributions.
The $\beta$-function then has the $\mu$-dependence one would expect in a
5-dimensional theory:
\beq
\frac{d\alpha^{-1}}{d\log\mu}=-\frac{1}{2\pi}\frac{\mu}{\mr}\trp Q^2\ ,
\eeq
where $\trp$ represents a sum over {\em only the massless states in the bulk}
but not their KK excitations. (The states confined to the walls are
also not included in the $\trp Q^2$ as they contribute only to the
logarithmic running which we are dropping.)
Running this down from $\mstrn$, one finds:
\beq
\alpha^{-1}(\mu)=\alpha^{-1}(\mstrn)+\frac{1}{2\pi}\frac{(\mstrn-\mu)}
{\mr}\,\trp Q^2
\label{rge}
\eeq
for $\mr\leq\mu\leq\mstrn$.
This running is linear in $\mu$, not logarithmic, and so $\alpha$ falls
rapidly. At scales $\mu\ll\mstrn$, Eq.~(\ref{rge}) reduces to:
\beq
\alpha(\mu)\simeq \left(\frac{2\pi}{\trp Q^2}\right)\frac{\mr}{\mstrn}\ .
\label{fp}
\eeq
independent of $\mu$;
even for $\trp Q^2$ as small as 1, $\alpha(\mu)$
is suppressed far below its string-scale value. In
fact, Eq.~(\ref{fp}) can be thought of as a quasi-fixed point solution to the
renormalization group equation~(\ref{rge}). 
For almost any initial value of $\alpha(\mstrn)$, the
value at $\mu\ll\mstrn$ will be given by Eq.~(\ref{fp}). The only
exception would be if $\alpha(\mstrn)\lsim(2\pi\mr)/(\trp
Q^2\mstrn)$. In the case at hand, with $\mstrn/\mr\simeq10$, the
validity and utility of this quasi-fixed point depends strongly on the value of
$\trp Q^2$.

We see that the strong running of $\alpha(\mu)$, caused by 
the extra dimension, can sharply reduce its value during inflation.
This strong running is however not experienced by 
$V_0= \frac12g^2\xi^2$.
By imposing the requirement $d V/d\mu=0$ at the scale $\mu=\phi$,
one finds the renormalization group equation
\beq
\frac{d \log V_0}{d \log\mu} = \frac{C\alpha(\mu)}{2\pi} \,.
\label{vrun}
\eeq
Taking $\alpha(\mu)$ from Eq.~(\ref{fp}), 
one sees that the fractional change in $V_0$ between $\mu=\mstrn$ and
$\mu=M_R$ is small. In contrast with the situation for $\alpha$,
it can be neglected just as is done for the case of $D$-term inflation
without large extra dimensions.\footnote
{As is usual, we are taking $\mu$ to have a fixed value, chosen
to be within the regime of $\phi$ where inflation takes place so
that two-loop and higher corrections are negligible. An alternative
procedure is to set $\mu=\phi$ and ignore all  loop corrections,
giving what is called 
the renormalization group improved tree-level potential.
The two procedures are equivalent provided that the variation of
$\phi$ during inflation is not too big, which is the case for
the present model. Indeed, one finds in that case that the
second procedure gives a potential that is practically linear in $\log\phi$,
in agreement with the first procedure.}

The strong damping of $\alpha$ due to the extra KK modes in the bulk
(for large $\trp Q^2$) 
gives us hope of naturally suppressing $\epsilon$ and $\eta$ so that inflation
can begin at $\phi_i\ll M_{11}$. Recall the self-consistency condition we derived
in Eq.~(\ref{self}). Plugging in the quasi-fixed point result for $\alpha$ produces:
\beq
M_{11}^2\gg\frac{C\alpha N}{\pi}\mpl^2\simeq\frac{2CN}{\trp
Q^2}\frac{\mr\mpl^2}{\mstrn}\simeq \frac{CN}{\trp Q^2}(10^{18}\gev)^2
\label{self2}
\eeq
which can only hold if
\beq
\frac{CN}{\trp Q^2}\ll 2\times10^{-3}
\label{self3}
\eeq
which in turn requires $\trp Q^2\gg10^4$! An equivalent statement is that
although the small value of $\alpha$ did indeed lower $\phi_i$, the extra
dimension also lowered the cut-off scale for the theory, by roughly the same
amount.
Thus, the HW-type models with one bulk dimension cannot solve the slow-roll
problem of 4-dimensional D-term inflation, unless the bulk is full of new
fields carrying the U(1) charge, enough to get $\trp Q^2\gg10^4$ (or,
alternatively, $\alpha(\mstrn)$ is far below its expected value).

The preceding analysis seems to give hope that in string theories with
more than one ``large'' dimension the fast running of $\alpha$ (which
goes as a power of the number of bulk dimensions) could solve the
problem of Planckian field values.
Consider what would happen if a string theory in $d$ dimensions compactified down to
$4+\delta$ dimensions at some string scale which, for simplicity, we
will also identify with the true $d$-dimensional Planck scale, $\md$.
Then at some second scale $\mr$ the remaining $\delta$ dimensions
compactify down to give the usual 4-dimensional world. Then we have
the relation $\md^{\delta+2}=\mr^\delta\mpl^2$.
The generalization of Eq.~(\ref{rge}) requires us to
count all the KK modes associated with the $\delta$-dimensions. In this case,
the KK states are labelled by a $\delta$-dimensional set of integers:
$m^2=\sum_{i=1}^\delta n_i^2\mr^2$.
However, now
the number of states with mass $m<\mu$  is given by $X_\delta
(\mu/\mr)^\delta$ where $X_\delta$ is the volume inside a
$\delta$-dimensional unit sphere. The renormalization group equation is then:
\beq
\frac{d\alpha^{-1}}{d\log\mu}=-\frac{1}{2\pi}\,X_\delta(\trp Q^2)
\left(\frac{\mu}{\mr}\right)^\delta
\eeq
for $\mr<\mu<\mn$. This too has a quasi-fixed point solution for
$\mu\ll\mn$, namely:
\beq
\alpha(\mu)\simeq \left(\frac{2\pi}{X_\delta\trp Q^2}\right)
\left(\frac{\mr}{\mn}\right)^\delta\ .
\eeq
Eqs.~(\ref{self}) and (\ref{self2}) generalize to 
\beq
\mn^2\gg\frac{2CN}{X_\delta\trp Q^2}\left(\frac{\mr}{\mn}\right)^\delta\mpl^2=
\frac{2CN}{X_\delta\trp Q^2}\,\mn^2
\eeq
for 
\beq
\frac{2CN}{X_\delta\trp Q^2}\ll 1\ .
\eeq
So we find that although the presence of extra dimensions could result
in a strong suppression of $\alpha$ as measured at the scale of the
vacuum energy, the corresponding decrease in the $d$-dimensional
Planck scale cancels the effect completely, so there is nothing to be
gained by increasing the number of ``large'' dimensions.

\section{Conclusions}

Inflationary models driven by non-zero Fayet-Iliopoulos terms, $\xi$,
suffer from a
slow-roll problem which requires inflation to begin when the inflaton field
value is Planckian. Furthermore, 
attempts to connect the source of the $\xi$-term to the
anomalous U(1) of string theory are inconsistent with the COBE data unless the
underlying string theory has some new scales present. The most promising class
of such models are the HW models which can be shown to produce $\xi$-terms
consistent with COBE. One might also expect these models to solve the slow-roll
problem (by suppressing $\alpha$)
since the U(1) gauge coupling has a strongly attractive quasi-fixed
point at very weak coupling if there is charged matter in the bulk.
However, the slow-roll problem remains even in these new models due to a
cancellation between the suppression of the gauge coupling and the suppression
of the $d$-dimensional Planck scale. Only in models with very large amounts of
matter in the bulk or strongly suppressed values of $\alpha(\mstrn)$
can the field theory be made self-consistent without
imposing stringent symmetries on the superpotential.

\section*{Acknowledgments}

We would like to thank Tony Gherghetta for discussions at the beginning of this
project. CK would also like to thank the Department of Physics at Lancaster University
for their hospitality while this work was completed.
This work was supported in part by NATO grant CRG--970214.
CK is also supported by US 
Department of Energy contract DE--AC03--76SF00098.


\begin{thebibliography}{99}

\bibitem{linde}
  A.~Linde, \PRD{49}{94}{748}.

\bibitem{lr}
  D.H.~Lyth and A.~Riotto, {\tt hep-ph/9807278}, to appear in Physics Reports.

\bibitem{finfl}
  E.~Copeland {\it et al.}, \PRD{49}{94}{6410}; \\
  M.~Dine, L.~Randall and S.~Thomas, \NPB{458}{96}{291}.

\bibitem{dinfl}
  E.~Stewart, \PRD{51}{95}{6847}; \\
  E.~Halyo, \PLB{387}{96}{43}; \\
  P.~Binetruy and G.~Dvali, \PLB{388}{96}{241}.

\bibitem{dsw}
  M.~Dine, N.~Seiberg and E.~Witten, \NPB{28}{87}{589}; \\
  J.~Atick, L.~Dixon and A.~Sen, \NPB{292}{87}{109}.

\bibitem{lrd}
  D.H.~Lyth and A.~Riotto, \PLB{412}{97}{28}.

\bibitem{jmr}
  J.~March-Russell, \PLB{437}{98}{318}.

\bibitem{hw}
  P.~Ho\v rava and E.~Witten, \NPB{460}{96}{506}; \NPB{475}{96}{94}.

\bibitem{bddp}
  P.~Binetruy, C.~Deffayet, E.~Dudas and P.~Ramond, {\tt hep-th/9807079}.

\bibitem{lpt}
  Z.~Lalak, S.~Pokorski and S.~Thomas, {\tt hep-ph/9807503}.

\bibitem{km}
  C.~Kolda and J.~March-Russell, {\tt hep-ph/9802358}.

\bibitem{gcu}
  E.~Witten, \NPB{471}{96}{135}.

\bibitem{bd}
 T.~Banks and M.~Dine, \NPB{479}{96}{173}.

\bibitem{lyth}
  D.H.~Lyth, {\tt hep-ph/9810320}.

\bibitem{ddg}
  K.~Dienes, E.~Dudas and T.~Gherghetta, \PLB{436}{98}{55}; 
  {\tt hep-ph/9806292}.

\end{thebibliography}
\end{document}